# Dispersion of third-harmonic generation in ultrastrongly coupled organic cavity polaritons

Bin Liu[1,†,*,] Michael Crescimanno[2], Robert J. Twieg[3], Kenneth D. Singer[1*]

[1]Department of Physics, Case Western Reserve University, Cleveland, Ohio 44106, USA

[2]Department of Physics and Astronomy, Youngstown State University,
Youngstown, Ohio 55444, USA

[3]Department of Chemistry and Biochemistry, Kent State University, Kent, OH 44242, USA

*Email: bxl224@case.edu, kds4@case.edu

**Abstract:** Organic cavity polaritons are bosonic quasi-particles that arise from the strong interaction between organic molecular excitons and photons within microcavities. The spectral dispersion of third harmonic generation near resonance with the cavity polariton states is studied experimentally via angle-resolved reflected third harmonic generation measurements with several pump wavelengths. In addition, a three-step nonlinear optical transfer matrix model is used to simulate the third harmonic generation using the sum-over-states dispersive nonlinear coefficients, which include hybrid exciton-photon polariton states. The angle-dependent experiment and modeling agree, revealing that the output of third harmonic generation is resonantly enhanced when the third harmonic wavelength is near the upper polariton state. The degree of enhancement is higher on the exciton-like branch of the polariton dispersion. Lower polariton enhanced third harmonic generation is not experimentally observed due to inadequate coupling at the longer wavelength, which is also consistent with the nonlinear transfer matrix modeling. These results indicate that the sum-over-states nonlinear dispersion is descriptive of the process implying that the polariton states are terms in a complete set of states forming the basis for the perturbative calculation of the nonlinear optical response. The results also indicate the possibility of wavelength agile nonlinear optical response by angle-of-incidence tuning.

Strong exciton-photon coupling within microcavities results in new eigenstates of microcavity polariton systems, which are characterized by anti-crossing of the momentum-energy dispersions of the cavity and exciton.[1] Due to the large oscillator strength and binding energy of organic materials, organic cavity polaritons exhibit large vacuum Rabi splitting energies ($\hbar\Omega_R$) even at room temperature.[2-4] When $\hbar\Omega_R$ is comparable (20%) to the uncoupled exciton energy ($E_{ex}$), the system is characterized by the ultrastrong coupling (USC) regime, where the rotating wave approximation (RWA) is no longer applicable, and the antiresonant terms can significantly modify cavity-polariton properties.[5-8]

The hybrid light-matter characteristic of both single and multiple cavity polaritons are of significant theoretical and practical importance, such as organic LEDs,[9,10] room-temperature polariton condensation and lasing,[11-13] polariton-mediated energy transfer,[14-16] conductivity enhancement,[17,18] and superfluidity.[19] However, nonlinear optical processes, such as the optical Kerr effect and third harmonic generation, arising from cavity polaritons in organic materials are not well studied. Recently, the enhancement of resonant second-harmonic generation (SHG) of the lower polariton is reported from strongly coupled organic crystalline nanofiber microcavities, where the SHG wavelength is resonant with the lower polariton.[20] Also, the enhanced third-harmonic generation (THG) is observed from cavity polaritons in the ultrastrong coupling regime when the pump wavelength is resonant with the lower polariton.[21]

Here, we experimentally and theoretically demonstrate upper-polariton enhanced THG using angle-varying reflection configuration in the ultrastrong coupling regime, and observe the THG intensity is largest when the THG wavelength is resonant with the upper polariton, in contrast to the observation of Barachati et al..[21] Our experiments also reveal that the enhanced THG is stronger when THG energy (wavelength) is resonant with the exciton-like upper cavity polariton branch.

Below we detail significant semi-quantitative agreement between these data and results from nonlinear transfer matrix modeling. That modeling incorporates dispersion in the underlying THG coefficients $\chi^{(3)}$ and allows us to test the applicability of the sum-over-states dispersion model for the THG coefficients with the polariton states dominating the state sum. A simplified quantum Hamiltonian approach is developed describing how the cavity coupling shifts the poles in the $\chi^{(3)}$ from excitonic to polaritonic. This profound change in the $\chi^{(3)}$ should generically enhance THG at both polariton resonances, and imprint the THG's angle dependence. We look for both effects and use the theory developed to understand the angle and wavelength dependence observed experimentally.

**RESULTS AND DISCUSSION**

The cavity structure, shown in the inset of **Figure 1**, is fabricated using metal mirrors enclosing a neat organic dye glass, DCDHF-6-V, which is in the push-pull DCDHF class previously used in a variety of optoelectronic experiments.[22,23] The complex refractive indices of the organic dye and its molecular structure are shown in **Figure 1**. The complex refractive indices are obtained by detailed modeling of spectroscopic ellipsometry.[24,25] To characterize the linear optical properties, the angularly resolved *s*-polarized reflectivity spectra of a 165 nm-thick organic microcavity are measured at room temperature (**Figure 2**a). The dispersion of cavity polaritons is characterized by the anti-crossing, yielding a vacuum Rabi splitting of $\hbar\Omega_R$ =1.12 eV (**Figure S2**a), which is about 52% of the uncoupled exciton energy, thus indicating ultrastrong exciton-photon coupling in the microcavity.[8] The dispersion of the cavity polariton calculated using the linear transfer matrix formalism are shown by the black curves in **Figure 2**a, which are in excellent quantitative agreement with the measured results.

Next, the angle-resolved *s*-polarized THG measurements are performed with the ultrastrongly coupled microcavity pumped at various fundamental wavelengths (see **Methods and Experimental** and **Supporting Information** for more details). **Figure 3**a shows the measured THG results for the 165 nm-thick cavity. For each pump wavelength, the THG is measured by varying the incident angle from 15º to 60º, and also from 60º to 15º, indicating the reproducibility of the measurements. Importantly, to verify the detected signals arise from THG, the peak THG intensity for various pump wavelengths at particular angles is measured as a function of the pump power (**Figure 3**c), which fit well to a cubic-power function, indicating the key characteristic of THG. As shown in **Figure 3**a, the THG intensity has a peak at a particular incident angle for each pump wavelength, and the angular position of the peak THG intensity shifts when the pump wavelength varies. Furthermore, for a fixed pump intensity, the longer (lower) the pump wavelength (energy), the larger the peak THG intensity. Thus, the angle values of the peak THG intensity for different pump wavelengths are plotted with the cavity-polariton dispersions in **Figure 3**b, and two different areas of the microcavity sample are tested. Also, the variation of the THG intensity is indicated by the change of the symbols' color from violet to red as indicated in the caption of **Figure 3**b. Two important observations should be emphasized. First, the THG energy (wavelength) with the angle of the peak THG intensity for different pump wavelengths follows the dispersion of the upper polaritons, which is the strong evidence of resonant cavity-polariton enhancement effect for THG. Second, the THG intensity is larger when the pump energy (wavelength) is lower (longer), indicating that resonant enhancement is larger when the exciton-like branch of the upper polariton state is resonant with the THG energy (wavelength). This observation implies that the nonlinear characteristic of organic dye molecules in the cavity-polariton state plays a more significant role in the THG intensity for the hybrid light-matter system.

To investigate possible thickness-related artifacts, a 140 nm-thick organic microcavity is also studied using both angularly resolved reflectivity measurements in the linear optical regime and angle-resolved THG measurements. **Figure 2**b shows the anti-crossing dispersion of the cavity polaritons, which also yields a vacuum Rabi splitting of $\hbar\Omega_R$ =1.12 eV (**Figure S2**b). **Figure 3**d shows the angle-resolved *s*-polarized THG measured results for the 140 nm-thick cavity, and the angle values of the peak THG intensity for different pump wavelengths are displayed in **Figure 3**e, which again follow the dispersion of the upper-polariton branch, suggesting the cavity-polariton enhanced THG. Moreover, considering the fixed pump intensity, the THG intensity is larger when the cavity-polariton state resonant with the THG wavelength is more exciton-like.

**THEORETICAL MODELING**

*1. Background*

The theoretical model we describe below is a novel extension for the cases of strong and ultrastrong coupling to the existing theoretical framework for THG in complex optical geometries.[26,27] To motivate this extension, consider a simplified microphysical model consisting of linear string of *N* identical two-level molecules (parameterized by a single 'exciton' energy interval, $\hbar\omega_e$, and dipole matrix element, $\mu$, for the isolated molecule), labelled by *j*=1,…, *N*, laid out periodically along the depth of a cavity (layer thickness *L*) having a Fabry-Perot resonance $\hbar\omega_c$ in the absence of any coupling near $\hbar\omega_e$. We make no a-priori assignment of definite parity to the molecular states connected by the dipole matrix element $\mu$ as is consistent with the non-centrosymmetric nature of the DCDHF-6-V chromophore. The truncated Hamiltonian of the cavity-molecules coupled system is thus,

$$H = \hbar\omega_c a^\dagger a + \hbar\omega_e b_k^\dagger b_k + i\hbar\mu\sqrt{N}(a^\dagger - a)(b_k^\dagger + b_k) + \frac{\hbar\mu^2 N}{f\omega_c}(b_k^\dagger + b_k)^2 + \cdots \quad (1)$$

where we have reorganized the molecular response operators into the sum $b_k = \frac{1}{\sqrt{N}}\sum_{j=1}^{N} b_j e^{ijkL}$, the only linear combination of molecular operators that have significant net non-zero overlap dipole matrix elements with the cavity mode of frequency $\omega_c$. The $f$ is the square of the 'filling factor' which we have included in the exciton-exciton interaction term. Note that the label '$k$' is not an integer, but rather a combination of the wave vector of the cavity mode and the interparticle spacing. The $b_k$ so defined inherits the canonical commutation relations of the excitonic operators $b_j$. The factor of $\sqrt{N}$ in front of the off-diagonal matrix elements of the Hamiltonian in Eq. (1) comes from the fact that all the molecules have an identical dipole matrix element, $\mu$. Thus, a single linear combination of dipoles with nonzero overlap with that cavity mode contributes. This leading factor of $\sqrt{N}$ causes the polariton splitting at degeneracy ($\omega_c=\omega_e$) to be proportional to $\sqrt{N}$, consistent with the observed chromophore density dependence. The ellipses in Eq. (1) indicate other unlisted terms that, although explicitly time-periodic in the RWA frame implied in Eq.(1), contribute significant lineshape shifts and asymmetries due to ultrastrong coupling.

We now perform an additional unitary transformation $U$ into the polariton basis as described,[28-30]

$$\left(P_{LP}, P_{UP}, P_{LP}^\dagger, P_{UP}^\dagger\right)^t = U(a, b, a^\dagger, b^\dagger)^t \quad (2)$$

which conveniently diagonalizes the Hamiltonian while retaining the canonical commutation relations. The polariton states are orthogonal mixed cavity-exciton states, and are the convenient basis from which to construct a perturbative theory of the system's linear and nonlinear optical response. The resulting energy eigenvalues are to the leading order (and for simplicity at large $f$):

$$\omega_{UP,LP} = \frac{1}{2}\left(\omega_e + \omega_c \pm \sqrt{(\omega_e - \omega_c)^2 + 16\mu^2 N}\right) \quad (3)$$

Including the THG process from a subharmonic pump at frequency $\omega$ can be done by taking into account the off-resonant polarizability of the system. Quantum optical evaluation of the THG process indicates that the net THG coefficient (the coefficient relating the cube of the local pump field at frequency $\omega$ with the THG output field at $3\omega$) has poles where $\omega$, $2\omega$ and $3\omega$ coincide with the excited states of the system, in this case, $\omega_{(UP,LP)}$, the polariton energies, and not at the (original) exciton or cavity frequencies. The resulting THG coefficient is a version of the sum-over-states approach,[31,32]

$$\chi^{(3)}(\omega) = \frac{\mu_{01}^2}{4\hbar^2}(\Delta\mu_{10}^2 D_{111} - \mu_{01}^2 D_{11} + \sum_n \mu_{1n}^2 D_{1n1}) \qquad (4)$$

where $D_{111}$ (the mixed parity contribution) has poles at the energy interval given by the {$\omega$, $2\omega$, $3\omega$}={$\omega_{LP}$, $\omega_{UP}$}, $D_{11}$ (the fixed parity multiphoton contribution) has poles only at {$\omega$, $3\omega$ }={$\omega_{LP}$, $\omega_{UP}$}, and $D_{1n1}$ (the excited state's contributions) has poles where these low harmonics of the pump would be degenerate with higher excited states (the 'off-resonant' AC polarizability term). Each of the $D$'s in Eq. (4) have the form of products of Lorentzian (simple poles in the complex plane), and thus also depend on the width of the contributing states. In the expression above, the $\mu_{ij}$ are products of dipole matrix elements of the associated states.

In the present application we use Eq. (4) as a phenomenological input to the nonlinear transfer matrix approach we describe below to simulate the expected THG signal. This is appropriate since relatively little is known in detail about the complex level structure and matrix elements of this dye molecule.

 *2. Transfer Matrix Methodology*

The method of nonlinear transfer matrices is used to calculate the reflected THG from the cavity-polariton system. This method conveniently analyzes the nonlinear response of optical harmonic generation from a multilayer structure consisting of any number of parallel slabs of arbitrary thickness.[26,27] Throughout the analysis below we work in the undepleted pump limit.

First, due to the fact that the equations governing the propagation of the fields are linear and that the tangential component of the electric field is continuous, multilayer structures with isotropic and homogeneous media and parallel-plane interfaces can be described by 2 x 2 matrices,[33,34] associated with a single polarization (conventionally *s* or *p*, however in what follows here we describe the method for *s*-polarization only), thus, the optical electrical field can be resolved into two components corresponding to the resultant total electric field; one component propagating in the positive direction, $\mathbf{E}^{(+)}$ and one in the negative direction, $\mathbf{E}^{(-)}$, respectively. This is true of both the pump field and the THG field.

Once the spatial distribution of the pump field is known the relevant polarization for THG is computed by cubing the pump field and multiplying it by the third-order susceptibility tensor, $\chi^{(3)}$, as,

$$\mathbf{P}^{3\omega} = \chi^{(3)} : \left(\mathbf{E}^{(+)} \exp(ik_z^\omega z) + \mathbf{E}^{(-)} \exp(-ik_z^\omega z)\right)^3$$
$$= \mathbf{P}^{I+} \exp(ik_z^{S,I} z) + \mathbf{P}^{I-} \exp(-ik_z^{S,I} z) + \mathbf{P}^{II+} \exp(ik_z^{S,II} z) + \mathbf{P}^{II-} \exp(-ik_z^{S,II} z) \quad (5)$$

where $k_z^\omega = k_\omega \cos\theta$ is wave vector inside the organic layer projected onto the normal to the interface. All cubic polarization terms are divided into two types. The terms of type I have the wave vector component $k_z^{S,I} = 3k_z^\omega$ corresponding to addition of three wave vectors propagating in the same direction, and the type II terms have the normal projection of the wave vector $k_z^{S,II} = k_z^\omega$ corresponding to the addition of three waves where one is counterpropagating relative to the other two.

The equation of inhomogeneous third-harmonic waves for direct THG $\chi^{(3)}$ is described as following,[26,27]

$$\mathbf{E}^{(s)I,II} = \frac{4\pi}{\epsilon^{I,II} - \epsilon(3\omega)} \left(\mathbf{P}_y^{I,II} + \mathbf{P}_\perp^{I,II}\right) - \frac{4\pi}{\epsilon(3\omega)} \mathbf{P}_\parallel^{I,II} \quad (6)$$

where $\epsilon^{I}=\epsilon(\omega)$ and $\epsilon^{II}=\epsilon(\omega)(1+8\sin\theta)/9$. Here, the cubic polarization has been conveniently decomposed into components with polarization directions normal ($\mathbf{P}_\perp$) and parallel ($\mathbf{P}_\parallel$) to the wave vector of the inhomogeneous wave, and the *s*-polarized component, $\mathbf{P}_y$.

The amorphous DCDHF-6-V layer can be treated as consistent with the isotropic point group with 21 nonzero elements for the $\chi^{(3)}$ tensor, of which only 3 are independent.[35] They are:

$$\begin{aligned}
\chi_{yyzz} &= \chi_{zzyy} = \chi_{zzxx} = \chi_{xxzz} = \chi_{xxyy} = \chi_{yyxx}, \\
\chi_{yzyz} &= \chi_{zyzy} = \chi_{zxzx} = \chi_{xzxz} = \chi_{xyxy} = \chi_{yxyx}, \\
\chi_{yzzy} &= \chi_{zyyz} = \chi_{zxxz} = \chi_{xzzx} = \chi_{xyyx} = \chi_{yxxy}, \\
\chi_{xxxx} &= \chi_{yyyy} = \chi_{zzzz} = \chi_{xxyy} + \chi_{xyxy} + \chi_{xyyx}
\end{aligned} \qquad (7)$$

For *s*-polarized fundamental wavelength, the type I and type II cubic polarization are determined after some algebra, which gives

$$\mathbf{P}_y^I = \frac{1}{3}\chi E_y^3, \quad \mathbf{P}_y^{II} = \chi E_y^3. \qquad (8)$$

where only the *s*-polarized component is induced. $E_y$ is the *s*-polarized fundamental electrical field, and $\chi=\chi_{xxxx}$ is the only tensor element needed for this case with *s*-polarized pump.

The third-harmonic polarization, Eq. (6), enters as an inhomogeneous 'source' term in the Maxwell's equations for the THG field component. The transfer matrix derived in ref [26,27] solves this Maxwell equations with the appropriate boundary conditions for the THG field component (i.e. at $3\omega$), from which we find the third-harmonic field emitted into the air adjacent to the top Ag layer, which is the reflected THG, as well as the field into the air adjacent to sample substrate, which is the transmitted THG (which we do not measure in this experiment). Note that the THG components from type I and type II nonlinear polarization terms actually represent light components that emerge at different angles from the sample. The angular acceptance of the detector used in the experiments we report on below was narrow enough to only receive appreciable THG signal from the type I terms and so we have focused only on that component only in our modeling.

*3. Simulation Results*

Considering the measured linewidth of cavity polariton states, nonlinear transfer-matrix simulation yields the calculation results for the 165nm and 140nm cavity, which are shown in **Figure 4**a and **4**c, respectively. Comparing **Figure 3**a and **Figure 4**a, the calculated angle-resolved THG agrees well with the experimental data for each pump wavelength, especially the angle locations of peak THG intensity for different pump wavelengths as shown in **Figure 4**b and **4**d for the 165nm and 140nm cavity, respectively. It is also noted that the calculated enhanced THG is larger for longer pump wavelength, which again agrees with experimental observation that THG intensity is larger when the THG wavelength is closer to the uncoupled exciton wavelength, that is, on the exciton-like branch of the polariton dispersion.

*4. Discussion on LP Enhanced THG*

The lower polariton enhanced THG was not observed experimentally. Even though the reflection spectrum suggests that THG should be resonantly enhanced when $3\omega$ coincides with the lower polariton state, one must consider the magnitude of the pump electric field at frequency $\omega$ in the cavity subject to the degree of non-resonant confinement at that frequency. In **Figure 5**a, our modeling indicates that the intercavity pump field amplitude at these longer pump wavelengths is typically 3-4 times smaller than at the shorter pump wavelengths associated with the upper polariton enhanced THG, leading to a naively 30 to 60-fold reduction in the THG output on the LP.

Supporting this, introducing the cavity-polariton-dependent $\chi^{(3)}$, we perform nonlinear transfer-matrix simulations with a broad range of pump wavelengths for a particular angle, using both polariton state locations and linewidths as measured by experiment. The resulting simulated THG output is shown in **Figure 5**b. The significant THG peak occurs for a pump wavelength of 1392 nm, where the THG wavelength (464 nm) is exactly resonant with the upper polariton at 0º. A second THG peak is also resolved, which is induced by the lower polariton, but it roughly 60 times smaller than that enhanced by the upper polariton.

**CONCLUSIONS**

We performed experimental studies and theoretical modeling of angle-resolved THG in ultrastrongly coupled organic cavity polaritonic matter. We find agreement with the angle-dependence of the THG experimental data and our nonlinear transfer matrix model when using a sum-over-states dispersion for $\chi^{(3)}$ dominated by the cavity polariton resonances. In the zero linewidth limit the cavity polariton states form a complete set of basis states for perturbation theory used in computing higher harmonic response. We also confirm by calculation and non-detection that the THG enhancement on the lower polariton state is too small to measure, primarily due to the poor cavity-pump coupling at these longer wavelengths. Our work showing very large Rabi splittings in low-Q metal-based organic microcavities at room temperature could lead to promising optoelectronic applications, and the large nonlinear optical response of cavity polaritons could provide potential for fundamental and applied nonlinear optics. Further, wavelength agile devices are possible by angle tuning the polariton states.

**METHODS AND EXPERIMENTAL**

*Organic Microcavity Fabrication*: A 100 nm-thick Ag layer was thermally evaporated onto a 1 mm-thick glass substrate in vacuum at $10^{-7}$ Torr. Solutions of DCDHF-6-V in toluene were spin cast on top of the metal film, with the thickness of the organic film varied by adjusting the spin speed. Microcavity fabrication was completed by evaporating a second 30 nm-thick Ag layer on top of the organic film, which provided microcavities with low Q-factor around 20.

*Profilometry, Spectroscopic Ellipsometry and Linearly Optical Angle-Resolved Reflectivity Measurements*: DCDHF-6-V layer thickness was measured using a stylus profilometer (KLA-Tencor P-6). Spectroscopic ellipsometry measurements were carried out with a Horiba Jobin Yvon UVISEL iHR320 ellipsometer under incident angles of 55°, 60°, and 65° for photon energies between 1 eV and 5 eV with 10 meV increments, and the complex refractive index spectra of DCDHF-6-V were obtained by fitting the ellipsometric data using the Forouhi-Bloomer model ("New Amorphous").[24] The linearly optical angle-resolved reflectivity spectra of the organic microcavities were acquired at room temperature using a J. A. Woollam Co. Inc V-VASE instrument with an angular increment of 5° and wavelength spectroscopic resolution of 2 nm.

*Angle-Resolved THG Measurements*: The organic microcavities were pumped by fundamental tunable IR wavelengths generated from a traveling-wave optical parametric amplifier of superfluorescence (TOPAS, Light Conversion Ltd.), which was pumped by a Ti:Sapphire regenerative amplifier (CPA-2010, Clark-MXR) with 1 kHz reparation rate, 200 fs pulse duration and 1 mJ pulse energy. The incident angle was varied by rotating the microcavities between 15° and 60° with an angular resolution of 0.5°, and the reflected THG intensity was collected by a photomultiplier tube (PMT, R1333, Hamamatsu) with a collecting lens and optical filters, which were placed on the rotating arm of a goniometer (**Figure S2**). The collecting lens and PMT combination had an angular acceptance of less than +/- 15° for emission from the sample. See **Supporting Information** for more details.

**Corresponding Author**

\* Email: bxl224@case.edu, kds4@case.edu

**Present Addresses**

†Department of Physics, City College of New York, New York, NY 10031, USA

**Acknowledgements**

This work was supported by the U.S. National Science Foundation (Grant DMR-1609077). The authors acknowledge the use of the Materials for Opto/Electronics Research and Education Center (MORE) for sample preparation and characterization at Case Western Reserve University.

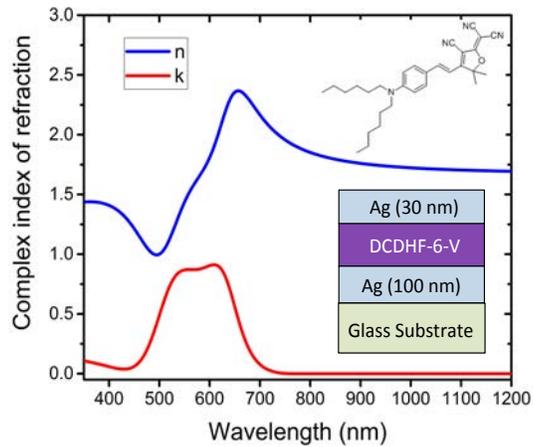

**Figure 1**. Complex refractive index of neat DCDHF-6-V film. The inset shows (top) the chemical structure of the organic dye molecule and (bottom) microcavity structure.

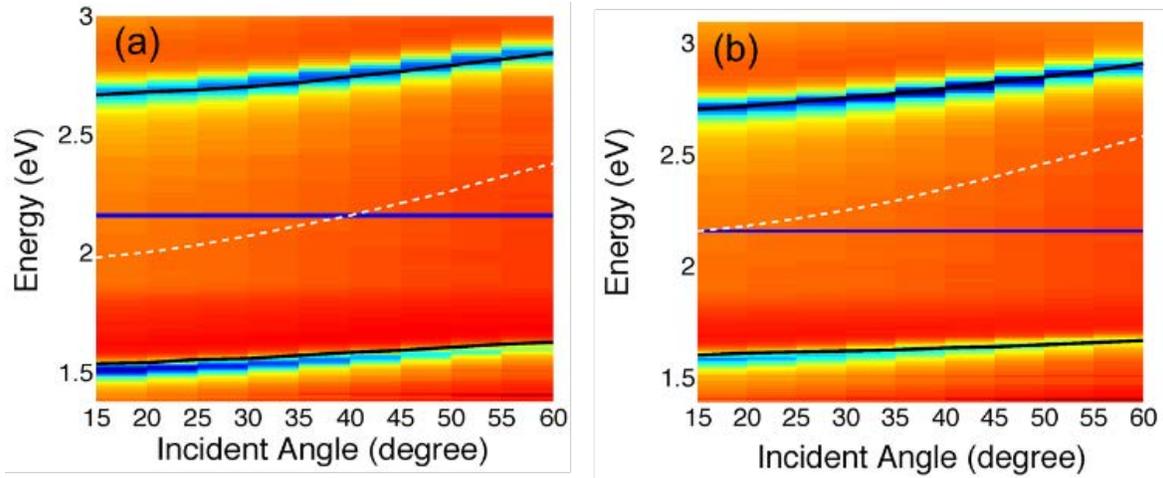

**Figure 2**. Angle-resolved *s*-polarized reflectivity map of a **(a)** 165 nm-thick, and **(b)** 140 nm-thick microcavity, respectively. The black curves show the calculated polariton dispersion using the linear transfer matrix method. The dashed white curve is the dispersion of bare cavity mode, and the blue line shows the DCDHF-6-V exciton energy.

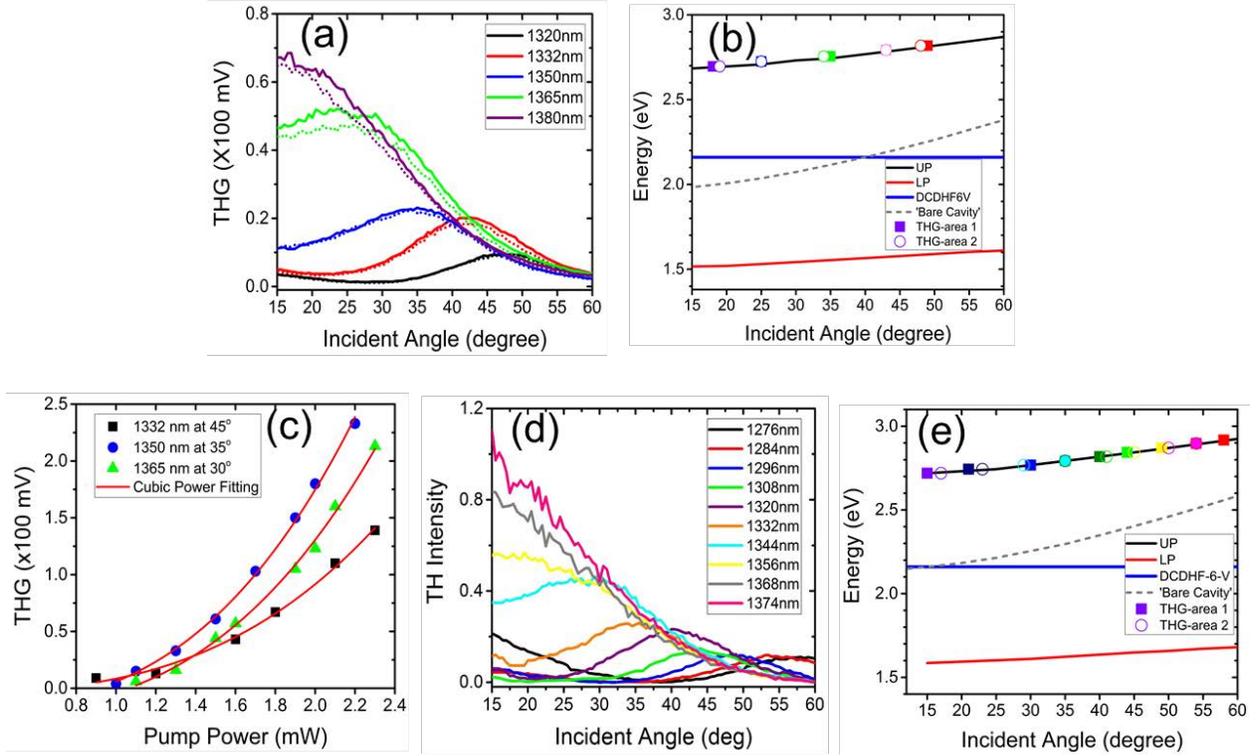

**Figure 3**. **(a)** Angle-resolved reflected THG intensity of the 165 nm-thick cavity pumped at different fundamental wavelengths. The solid curves show the results when the incident angle varies from 15º to 60º, and the dotted curves indicate the incident angle varies from 60º to 15º. The pump power is fixed as 1.3 ± 0.2 mW, with the diameter of the beam spot of 80 μm. **(b)** THG energy versus the incident angle at which the peak THG intensity is generated, following the dispersion of the upper-polariton branch. Solid squares show the results in **(a)**, and the circles represent the results from a different measurement area of the same microcavity. The scattered squares and circles with color changing from violet to red represent the THG intensity monotonically decreases. **(c)** Measured THG intensity as a function of pump power for different wavelengths, and the red curves indicate the cubic power fitting. **(d)** Angle-resolved reflected THG intensity of the 140 nm-thick cavity pumped by different fundamental wavelengths. **(e)** THG energy versus the incident angle at which the peak THG intensity is generated, following the dispersion of the upper-polariton branch. The squares and circles show the results from two

different areas of the same microcavity sample, and variation of the symbols' color from violet to red represent the THG intensity monotonically decreases.

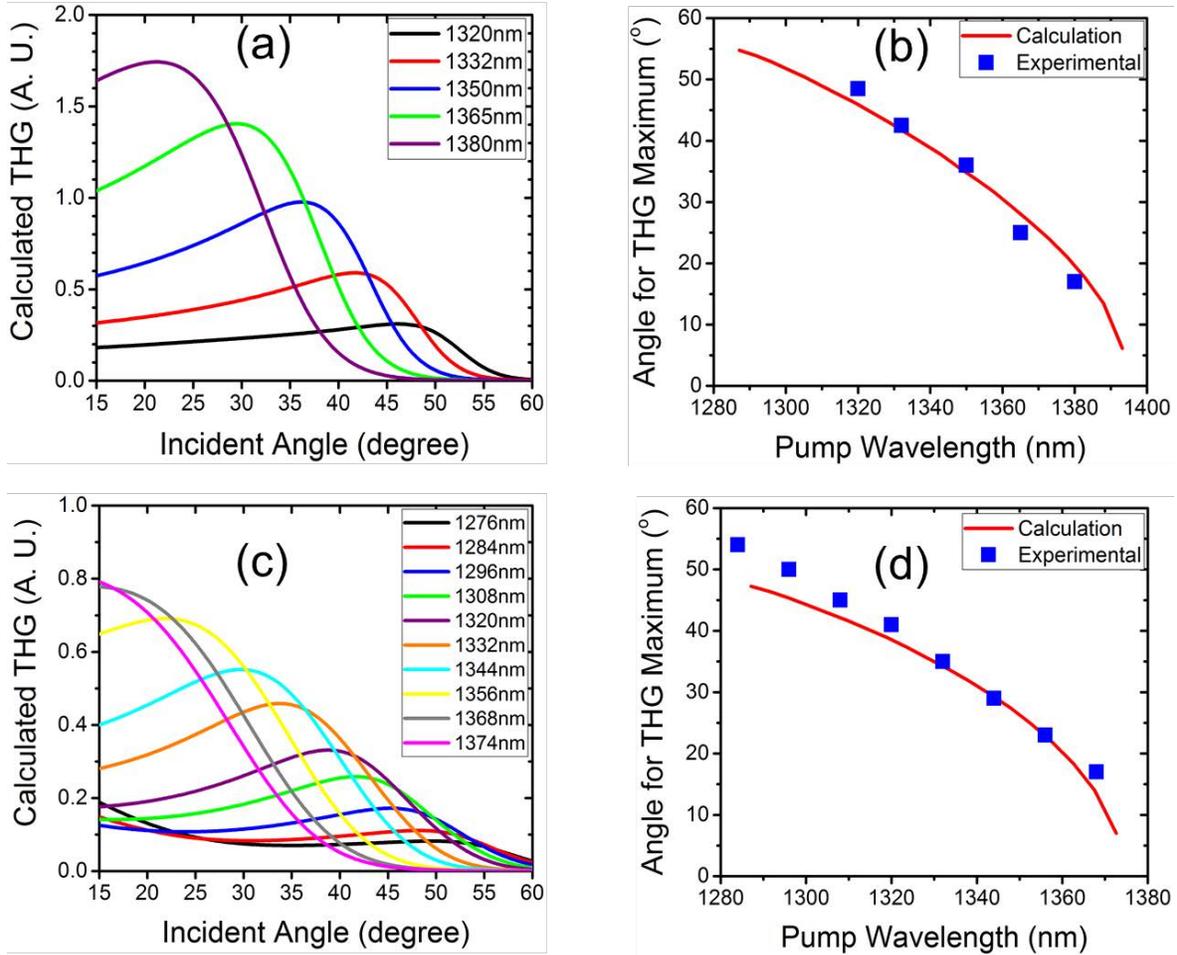

**Figure 4**. Calculated angle-resolved reflected THG of the **(a)** 165 nm, and **(c)** 140 nm-thick cavity. The comparison between calculation results and experimental data for the relation of THG peak angle versus the pump wavelength: **(b)** 165 nm cavity, **(d)** 140 nm cavity.

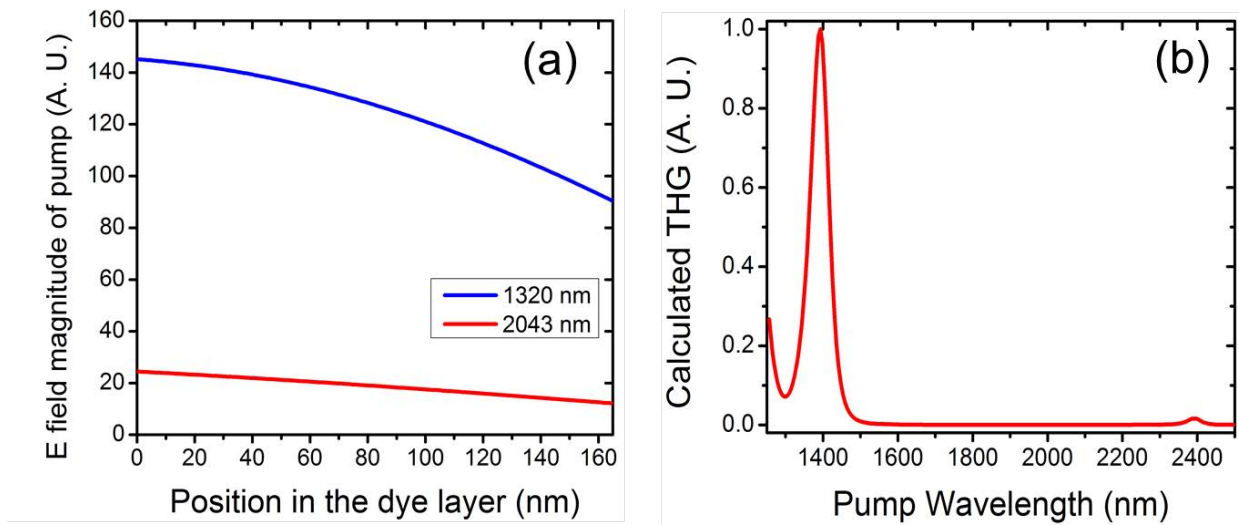

**Figure 5**. **(a)** Calculated field intensity distribution within the 165nm cavity for different pump wavelengths. **(b)** Calculated THG as a function of the pump wavelength for the 165nm cavity at the normal incidence.

Reference:

(1) Weisbuch, C.; Nishioka, M.; Ishikawa, A.; Arakawa, Y. Observation of the coupled exciton-photon mode splitting in a semiconductor quantum microcavity. *Phys. Rev. Lett.* **1992**, *69*, 3314-3317.

(2) Lidzey, D. G.; Bradley, D. D. C.; Skolnick, M. S.; Virgili, T.; Walker, S.; Whittaker, D. M. Strong exciton-photon coupling in an organic semiconductor microcavity. *Nature* **1998**, *395*, 53-55.

(3) Lidzey, D. G.; Bradley, D. D. C.; Virgili, T.; Armitage, A.; Skolnick, M. S.; Walker, S. Room temperature polariton emission from strongly coupled organic semiconductor microcavities. *Phys. Rev. Lett.* **1999**, *82*, 3316-3319.

(4) Holmes, R. J.; Forrest, S. R. Strong exciton-photon coupling in organic materials. *Org. Electron.* **2007**, *8*, 77-93.

(5) Todorov, Y.; Andrews, A. M.; Colombelli, R.; De Liberato, S.; Ciuti, C.; Klang, P.; Strasser, G.; Sirtori, C. Ultrastrong light-matter coupling regime with polariton dots. *Phys. Rev. Lett.* **2010**, *105*, 196402.

(6) Kéna-Cohen, S.; Maier, M. A.; Bradley, D. D. C. Ultrastrongly coupled exciton-polaritons in metal-clad organic semiconductor microcavities. *Adv. Opt. Mater.* **2013**, *1*, 827-833.

(7) Gambino, S.; Mazzeo, M.; Genco, A.; Di Stefano, O.; Savasta, S.; Patanè, S.; Ballarini, D.; Mangione, F.; Lerario, G.; Sanvitto, D.; Gigli, G. Exploring light-matter interaction phenomena under ultrastrong coupling regime. *ACS Photonics* **2014**, *1*(10), 1042-1048.

(8) Liu, B.; Rai, P.; Grezmak, J.; Twieg, R. J.; Singer, K. D. Coupling of exciton-polaritons in low-Q coupled microcavities beyond the rotating wave approximation. *Phys. Rev. B* **2015**, *92*, 155301.

(9) Gubbin, C. R.; Maier, M. A.; Kéna-Cohen, S. Low-voltage polariton electroluminescence from an ultrastrongly coupled organic light-emitting diode. *Appl. Phys. Lett.* **2014**, *104*, 233302.

(10) Mazzeo, M.; Genco, A.; Gambino, S.; Ballarini, D.; Mangione, F.; Di Stefano, O.; Patanè, S.; Savasta, S.; Sanvitto, D.; Gigli, G. Ultrastrong light-matter coupling in electrically doped microcavity organic light emitting diodes. *Appl. Phys. Lett.* **2014**, *104*, 233303.

(11) Kéna-Cohen, S.; Forrest, S. R. Room-temperature polariton lasing in an organic single-crystal microcavity. *Nat. Photonics.* **2010**, *4*, 371-375.

(12) Plumhof, J. D.; Stöferle, T.; Mai, L.; Scherf, U.; Mahrt, R. F. Room-temperature Bose–Einstein condensation of cavity exciton–polaritons in a polymer. *Nat. Mater.* **2014**, *13*, 247-252.

(13) Daskalakis, K. S.; Maier, M. A.; Kéna-Cohen, S. Nonlinear interactions in an organic polariton condensate. *Nat. Mater.* **2014**, *13*, 271-278.

(14) Coles, D. M.; Somaschi, N.; Michetti, P.; Clark, C.; Lagoudakis, P. G.; Savvidis, P. G.; Lidzey, D. G. Polariton-mediated energy transfer between organic dyes in a strongly coupled optical microcavity. *Nat. Mater.* **2014**, *13*, 712-719.

(15) Zhong, X.; Chervy, T.; Wang, S.; George, J.; Thomas, A.; Hutchison, J. A.; Devaux, E.; Genet, C.; Ebbesen, T. W. Non-radiative energy transfer mediated by hybrid light-matter states. *Angew. Chem. Int. Ed.* **2016**, *55*, 6202-6206.

(16) Zhong, X.; Chervy, T.; Zhang, L.; Thomas, A.; George, J.; Genet, C.; Hutchison, J. A.; Ebbesen, T. W. Energy transfer between spatially separated entangled molecules. *Angew. Chem. Int. Ed.* **2017**, *56*, 9034-9038.

(17) Feist, J.; Garcia-Vidal, F. J. Extraordinary exciton conductance induced by strong coupling. *Phys. Rev. Lett.* **2015**, *114*, 196402.

(18) Schachenmayer, J.; Genes, C.; Tignone, E.; Pupillo, G. Cavity-enhanced transport of excitons. *Phys. Rev. Lett.* **2015**, *114*, 196403.

(19) Lerario, G.; Fieramosca, A.; Barachati, F.; Ballarini, D.; Daskalakis, K. S.; Dominici, L.; De Giorgi, M.; Maier, M. A.; Gigli, G.; Kéna-Cohen, S.; Sanvitto, D. Room-temperature superfluidity in a polariton condensate. *Nat. Phys.* **2017**, *13*, 837-841.

(20) Chervy, T.; Xu, J.; Duan, Y.; Wang, C.; Mager, L.; Frerejean, M.; Münninghoff, J. A. W.; Tinnemans, P.; Hutchison, J. A.; Genet, C.; Rowan, A. E.; Rasing, T.; Ebbesen, T. W. High-efficiency second-harmonic generation from hybrid light-matter states. *Nano Lett.* **2016**, *16*, 7352-7356.

(21) Barachati, F.; Simon, J.; Getmanenko, Y. A.; Barlow, S.; Marder, S. R.; Kéna-Cohen, S. Tunable third-harmonic generation from polaritons in the ultrastrong coupling regime. *ACS Photonics* **2018**, *5*(1), 119-125.

(22) Ostroverkhova, O.; Gubler, U.; Wright, D.; Moerner, W. E.; He, M.; Twieg, R. J. Recent advances in understanding and development of photorefractive polymers and glasses. *Adv. Funct. Mater.* **2002**, *12*, 621-629.

(23) Ostroverkhova, O.; Moerner, W. E.; He, M.; Twieg, R. J. High-performance photorefractive organic glass with near-infrared sensitivity. *Appl. Phys. Lett.* **2003**, *82*, 3602-3604.

(24) Löper, P; Stuckelberger, M.; Niesen, B.; Werner, J.; Filipič, M.; Moon, S. J.; Yum, J. H.; Topič, M.; De Wolf, S.; Ballif, C. Complex refractive index spectra of $CH_3NH_3PbI_3$ Perovskite thin films determined by spectroscopic ellipsometry and spectrophotometry. *J. Phys. Chem. Lett.* **2015**, *6*, 66-71.

(25) Liu, B.; Soe, C. M. M.; Stoimpos, C. C.; Nie, W.; Tsai, H.; Lim, K.; Mohite, A. D.; Kanatzidis, M. D.; Marks, T. J.; Singer, K. D. Optical properties and modeling of 2D perovskite solar cells. *Solar RRL* **2017**, *1*, 1700062.

(26) Bethune, D. S. Optical harmonic generation and mixing in multilayer media: analysis using optical transfer matrix techniques. *J. Opt. Soc. Am. B* **1989**, *6*, 910-916.

(27) Martemyanov, M. G.; Dolgova, T. V.; Fedyanin A. A. Optical third-harmonic generation in one-dimensional photonic crystals and microcavities. *J. Exp. Theor. Phys.* **2004**, *98*, 463-477.

(28) Hopfield, J. J. Theory of the contribution of excitons to the complex dielectric constant of crystals. *Phys. Rev.* **1958**, *112*, 1555-1567.

(29) Ciuti, C.; Bastard, G.; Carusotto, I. Quantum vacuum properties of the intersubband cavity polariton field. *Phys. Rev. B* **2005**, *72*, 115303.

(30) Ciuti, C.; Carusotto, I. Input-output theory of cavities in the ultrastrong coupling regime: The case of time-independent cavity parameters. *Phys. Rev. A* **2006**, *74*, 033811

(31) Kuzyk, M. G.; Dirk, C. W. Effects of centrosymmetry on the nonresonant electronic third-order nonlinear optical susceptibility. *Phys. Rev. A* **1990**, *41*, 5098-5109.

(32) Andrews, J. H.; Khaydarov, J. D. V.; Singer, K. D.; Hull D. L.; Chuang, K. C. Characterization of excited states of centrosymmetric and noncentrosymmetric squaraines by third-harmonic spectral dispersion. *J. Opt. Soc. Am. B* **1995**, *12*, 2360-2371.

(33) Pettersson, L. A. A.; Roman, L. S.; Inganäs, O. Modeling photocurrent action spectra of photovoltaic devices based on organic thin films. *J. Appl. Phys.* **1999**, *86,* 487–496.

(34) Knittl, Z. *Optics of Thin Films*; Wiley, London, 1976.

(35) Boyd, R. W. *Nonlinear Optics*; Elsevier Science, 2008.